\documentclass[11pt]{article}
\usepackage{moriond,epsfig}

\def\be{\begin{equation}}
\def\ee{\end{equation}}
\def\bea{\begin{eqnarray}}
\def\eea{\end{eqnarray}}

\begin{document}
\vspace*{4cm}
\title{DARK ENERGY, MOND AND SUB-MILLIMETER TESTS OF GRAVITY}

\author{I. NAVARRO$^*$ and K. VAN ACOLEYEN$^{\dagger}$}

\address{$^*$DAMTP, University of Cambridge, CB3 0WA Cambridge, UK\\
$^{\dagger}$IPPP, University of Durham, DH1 3LR Durham, UK
}

\maketitle\abstracts{We consider modifications of General
Relativity obtained by adding the logarithm of some curvature
invariants to the Einstein-Hilbert action. These non-linear
actions can explain the late-time acceleration of the universe
giving an expansion history that differs from that of a pure cosmological
constant. We show that they also modify the Newtonian potential
below a fixed acceleration scale given by the late-time Hubble
constant times the speed of light. This is exactly what is
required in MOND, a phenomenological modification of the Newtonian
potential that is capable of explaining galactic rotation curves
without the need to introduce dark matter. We show that this kind
of modification also predicts short distance deviations of
Newton's law at the sub-mm scale and an anomalous shift in the
precession of the Moon's orbit around the Earth, both effects of a
size that is less than an order of magnitude below current
bounds.}

\section{Modified gravity: motivation}

The validity of General Relativity (GR) has been extensively
tested not only in the Solar System but also in other
astrophysical systems that involve stronger gravitational
fields\cite{Will:2005va}. And while there is a widespread belief
among theoretical physicists that the Einstein equation will have
further corrections that will be computable once a consistent
quantization of gravity is achieved, these corrections are
expected to modify the behaviour of the classical solutions only at
very short distances, unaccessible to present and possibly future
experiments. According to this picture there would be a big
``gravitational desert'' for curvatures much smaller than the Planck mass
where GR would
provide an accurate description of gravity.

On the other hand, cosmological and astrophysical observations are
widely inconsistent with GR if we consider as the source in the
energy-momentum tensor the visible matter in many systems. Since
we have no obvious theoretical motivation for expecting an
infrared modification of gravity, the first hypothesis one is led to is that there are extra sources in the energy-momentum
tensor that do not interact with photons. In this case we need two
such components: dark energy and dark matter. The simplest way to
incorporate these components into the theory is to assume the
existence of a small vacuum energy (or cosmological constant) of
magnitude $\Lambda_{vac}^4 \sim (10^{-3}eV)^4$ and an extra weakly
interacting massive particle that makes up most of the matter density
of the Universe. This particle is assumed to be cold (i.e.
non-relativistic) and can be naturally produced with the right
abundance in the early universe if it is associated with the
electro-weak scale. This is the working hypothesis for the
standard $\Lambda$CDM model. This model is capable of explaining the
features observed in the temperature fluctuations of the CMB and is
in general good agreement with other cosmological
probes\cite{Spergel:2006hy} (although one could mention some tension
between determinations of the power spectrum amplitude coming from
CMB and Lyman-$\alpha$ forest\cite{Seljak:2006bg}).

But the discrepancies between theory and experiment begin to grow
as we go to shorter length scales and we compare the simulations
of structure formation in CDM models with
observations\cite{Ostriker:2003qj}. The most obvious problems come
from the ``cuspy core'' in the central parts of halos or the
abundance of substructures produced in the simulations, none of
which seem compatible with observations. This has prompted the
consideration of other flavors of dark matter like warm or
self-interacting dark matter in order to reduce the halo
sub-structure and its density in the innermost part.
The problem with these alternatives is that they tend to delay
structure formation in the early universe, and it is not clear to
what extent such possibilities are compatible with the observed
constraints coming from evolved structure seen at high
redshifts\cite{Primack:2002th}. But despite of these problems, the
most serious problem that the dark matter paradigm faces is
probably to explain the strong correlation between the luminous
matter content and the dark matter density that is inferred by the
study of the rotation curves of galaxies. For spiral galaxies
these rotation curves imply that the visible matter content
completely determines the dynamics and predicts the detailed
rotation curve\cite{Persic:1995ru} even when the visible matter
makes only a very small fraction of the dynamical mass. These
correlations are not restricted to spiral galaxies and are
exemplified in several well-known empirical relations like the
Tully-Fisher or the Faber-Jackson laws. The degree of precision of
these correlations, compatible with zero intrinsic scatter, is hardly expected from a stochastic structure
formation scenario of the kind envisaged in dark matter models.
Remarkably, a simple modification of the Newtonian potential for
small accelerations can account surprisingly well for this
phenomenology\cite{Sanders:2002pf} (although the simplest fitting formula does seem to find
problems both for larger cluster scales\cite{Sanders:2002pf} and small satellites of
galaxies\cite{Zhao:2005xk}). This is the so-called MOND (for
MOdified Newtonian Dynamics), proposed by Milgrom\cite{Milgrom:1983ca} in
1983 which simply states that below a fixed
critical acceleration the force of gravity decays with an $1/r$
law instead of the Newtonian $1/r^2$. An interesting observation
is that the critical acceleration implied by the data, $a_0\sim
1.2\times 10^{-10}m/s^2$, is of the order of the Hubble constant times
the speed of light, which is determined by the dark energy
density. This coincidence strongly suggests, within a modified theory of
gravity, a link between both phenomenons. It is clear thus what
kind of properties we should look for in a modification of gravity
if it is to replace dark matter.

A relativistic theory with a MOND-like Newtonian limit that agrees
with Solar System tests has been proposed\cite{Bekenstein:2004ne}.
It is built by adding extra fields to the action
with particular couplings and supplementing them with constraints
by introducing also Lagrange multipliers. This theory has been shown to be
consistent with other cosmological observations with the help of
massive neutrinos\cite{Skordis:2005xk}, but the relation $a_0 \sim
cH_0$ remains unexplained. Here we will present a modification of
the Einstein-Hilbert action for the space-time metric
\cite{Navarro:2005ux} such that we get a modification of the
Newtonian potential below a fixed acceleration scale and where the
relation $a_0 \sim cH_0$ is naturally explained. Moreover we will
see that these kind of theories make predictions for
deviations (with respect to GR) in measurable quantities at the Solar System and laboratory levels. In fact we
will see that some of these effects should be on the edge of
detection, and this opens the door to the possibility of getting an
experimental validation in the laboratory of modified theories of gravity intended
to address cosmological and astrophysical phenomena.

\section{Modifying gravity below a fixed acceleration scale}

The actions we will be interested on are of the type
\be
S=\int \!\!d^4x\sqrt{-g}\frac{1}{16\pi
  G_N}\left\{R-\mu^{2}{\rm
Log}\left[f(R,Q-4P)\right]\right\}\,,\label{actionMOND} \ee where
 \be P \equiv R_{\mu\nu}R^{\mu\nu}\,\;\;
{\rm and}\;\;\; Q \equiv
R_{\mu\nu\lambda\rho}R^{\mu\nu\lambda\rho}
\ee
and $f$ is a function for which we will only assume that $f\rightarrow
0$ for $R_{\mu\nu\lambda}^{\sigma}\rightarrow
0$, and we can approximate $f\simeq Q/Q_0$ whenever $Q\gg R^2,P$.
Minkowski spacetime will not be a solution of the
theory but there will typically exist de Sitter solutions with
curvature $R\sim \mu^2$. We see then that if we want to explain the late time acceleration of
the Universe we have to take $\mu \sim H_0$, with $H_0$ being the
value of Hubble's constant today. But even if these theories could
explain the acceleration of the Universe, they raise several serious
questions that should be addressed before one could consider them as a
candidate to explain such acceleration. For instance, since the
equations of motion for the spacetime metric now contain up to fourth
order derivatives, one can worry
about the unwanted appearance of ghosts that would render the vacuum
unstable. Also one can expect that the extra propagating degrees of freedom introduced by
these higher derivatives would modify the Newtonian limit, and one
should check that it is modified in a manner compatible with
observation. For studying these issues it is convenient to discuss the
linearisation, or particle content, of these modified theories of gravity.

\subsection{Particle content of modified gravity}

In general, if we consider an action that is an arbitrary function of
the invariants $R$, $P$ and $Q$ we can expect eight propagating
degrees of freedom in vacuum\cite{Hindawi:1995cu}. These are
grouped as: two in a massless spin two particle, one in a scalar excitation and
five in a massive spin two ghost\footnote{We assume that the
  overall sign is such that
the  massless spin two graviton and the scalar are not ghosts.}. It is
easy to obtain the properties of these degrees of freedom by realizing
that at the bilinear level, the expansion over a maximally symmetric
spacetime of any action defined
through a
Lagrangian such as ${\cal L}=F(R,P,Q)$ is
the same\footnote{Notice that this equivalence only
  applies to the expansion of the action up to the $bilinear$ level.} as the expansion of\cite{Navarro:2005ux,Hindawi:1995cu}
\bea S=\int
\!\!d^4x\sqrt{-g}\frac{1}{16\pi
  G_N}\left[-\Lambda + \delta R
+\frac{1}{6m_0^2}R^2-\frac{1}{2m_2^2}C^{\mu\nu\lambda\sigma}C_{\mu\nu\lambda\sigma}\right]
\label{expand-action},\eea where $C_{\mu\nu\lambda\sigma}$ is the
Weyl tensor and the parameters appearing in these action can be
obtained as functions of $F(R,P,Q)$ and its derivatives with respect
to $R$, $P$ and $Q$ evaluated in the background\cite{}. $m_0$ is the
mass of the scalar and $m_2$ is the mass of the ghost. In particular
$m_2^{-2}=-\left<\partial_PF+4\partial_QF\right>_0$ so for functions
of the type $F(R,P,Q)=F(R,Q-4P)$, the ghost is absent. This is the
reason why we took this particular combination of $P$ and $Q$ in our original action
(\ref{actionMOND}). For this action we find that in vacuum the mass of the
scalar is given by $m_0 \sim H_0^2/\mu \sim \mu$, so the extra scalar
is almost massless. This would appear to rule out the
theory, since we know that at the Solar System level gravity is
mediated only by a massless spin two graviton. But a closer inspection
of the expansion reveals that this conclusion is not correct. The
reason is that for corrections to the Einstein-Hilbert action that
become important at small curvatures (but are negligible at large
curvatures), the expansion of the action in powers of the fluctuations
breaks down at a very small energy scale\cite{Navarro:2005da}. This means that for
actions of the type (\ref{actionMOND}) the
spherically symmetric solution found in the linearised approximation
can not be trusted at distances smaller than
$r_V \equiv \left(G_NM/\mu^3\right)^{1/4}$.
For a star like the Sun this distance is of the order of $10$ $kpc$,
many orders of magnitude larger than the Solar System. 

\subsection{Newtonian limit of modified gravity}

If the linearised
approximation is not valid, and the full non-linear equations are
difficult to solve, how can we proceed? We can get some insight on
the expected behaviour of the solutions by applying the following argument. If we
have an extra degree of freedom with mass $m_s$, we can expect that it
will only affect the solution whenever $r<m_s^{-1}$. For longer
distances the mass effectively decouples
it. We can then estimate the mass of the extra scalar particle in a
generic background by
applying the expression that we found when linearising in maximally
symmetric spacetimes. On a
Schwarzschild background we find that the mass depends on the distance
as $m_0^2 \sim Q/\mu^2 \sim (G_NM)^2/(r^6 \mu^2)$. The relation
$r<m_0^{-1}$ now turns into
\be
r > \left(\frac{G_NM}{\mu}\right)^{1/2}\equiv r_c.
\ee
So due to the dependence of the scalar mass on the distance we see
that  we can
expect a $long$ distance modification of gravity. Moreover this long distance
corresponds to a fixed Newtonian acceleration scale $a_0 \sim \mu$, precisely of the order of the late-time Hubble constant times the speed of light (that we are setting to $1$), as required in
MOND. Notice that for the Sun $r_c \sim 10^3 AU$, where $1AU$ is the
Sun-Earth distance.

One can check this result more rigorously applying an
approximation procedure that is complementary to the linearised
approximation for this type of theories\cite{Navarro:2005gh}. This alternative expansion is valid whenever the
extra term that one adds to the Einstein tensor produces only a small correction of
the GR background. What one
can do then is to take as the 0-th order solution the solution of GR. Then
compute the correction term to the GR equations evaluated in this
background and solve for the backreaction in the Einstein
tensor. One can iterate this process and provided that the backreaction produces only a small perturbation
in the original background, one can expect that the procedure will
produce a good approximation to an exact
solution if we iterate the process a sufficient number of times. Doing
this it was shown\cite{Navarro:2005ux} that the corrections to the Schwarzschild geometry
are small at small distances ($r\ll r_c$) and take the form
\bea
ds^2 & \simeq &
-\left[1-\frac{2G_NM}{r}\left(1+\frac{4}{3}\left(\frac{r}{r_c}\right)^4
 +  {\cal
O}\left(\left(\frac{r}{r_c}\right)^{8}\right)\right) \right]dt^2\nonumber\\
&& +\left[ 1-\frac{2G_NM}{r}\left(1-2\left(\frac{r}{r_c}\right)^4 +  {\cal
O}\left(\left(\frac{r}{r_c}\right)^{8}\right)\right)\right]^{-1}dr^2 + r^2 d\Omega_2^2.
\eea
From this expression it is clear that the modifications of the
gravitational field of the Sun at the Solar System level are very
small. But in the Solar System there are very stringent tests of GR that we
will have to face, the most precise coming
probably form the Lunar Laser Ranging experiment. Using this the
Moon-Earth distance is known with a precision of a centimeter. Any
anomalous precession is bound to be less than $2.4\times 10^{-11}$ radians
per revolution\cite{Dvali:2002vf}. Considering the correction to the gravitational potential
of the Earth given by the expression above one can estimate the
expected anomalous precession in radians per revolution as
\be
\frac{d}{dr}\left(r^2\frac{d}{dr}\left(\frac{\delta
  V}{rV_N}\right)\right) \simeq 16
\pi\left(\frac{r_{(Moon-Earth)}}{r_{c(Earth)}}\right)^4 \sim 10^{-12}
\ee 
which is just a factor of five below the current bound. So this theory
passes the tests coming from precision astrometrical measurements in
the Solar System, but what about the tests of gravity at smaller
scales? The $1/r$ form of the potential for gravity has been tested
down to scales of the order of 0.2 millimeters\cite{Hoyle:2004cw}. What kind of short
distance deviations, if any, can we expect
in our case? In order to get some intuition about the expected
scale where anomalies could appear for this kind of experiments we can
again estimate the mass of the scalar field in the dominant background
of a massive object of mass $M$, at a distance $r_0$ from its centre,
as $m_0 \sim G_NM/(r_0^3 \mu)$. If we take $M$ as the mass of the
Earth and $r_0$ as its radius we see that we do not expect
modifications in the Newtonian potential on the laboratory at
distances bigger than $m_0^{-1}|_{Earth} \sim 0.1 mm$. Since this is
just compatible with current bounds it is interesting to explore this
issue a bit further, and compute the first correction to the Newtonian
potential for a probe mass $m$ situated at a distance $r_0$ of a big
massive object of mass $M$. We will consider a coordinate
system $(t,x,y,z)$ where the masses are separated in the $z$ direction
so the 0-th order GR solution reads
(in the weak field limit, for small $r\equiv |{\bf x}| =
\sqrt{x^2+y^2+z^2}\ll r_0$)
\be
 ds^2=-(1+2\Phi^{(0)})dt^2+(1-2\Psi^{(0)})d{\bf x}^2
\ee
where 
\be \Phi^{(0)}\approx\Psi^{(0)}\approx-\frac{G_Nm}{r}-\frac{G_NM}{r_0}(1-\frac{z}{r_0}+\frac{3}{2}\frac{z^2}{r_0^2}-\frac{r^2}{2r_0^2})\,. \ee
We can now find the first order perturbation of the metric by solving
\be G_{\mu\nu}^{(1)}=-\mu^2 H_{\mu\nu}^{(0)} \,,\label{EOM1} \ee
where $\mu^2 H_{\mu\nu}$ is the extra term introduced in the equations
for the metric by the logarithmic part of the action. Taking
$g_{\mu\nu}=g_{\mu\nu}^{(0)}+g_{\mu\nu}^{(1)}$, the first order perturbation of the 00 and trace components of the
Einstein tensor are given by
\be G_{00}^{(1)}\approx 2\nabla^2\Psi^{(1)}\, ,\;\;\;\;
{G_\mu^\mu}^{(1)}\approx 2\nabla^2\Phi^{(1)}-4\nabla^2\Psi^{(1)}\,, \ee 
while for $\mu^2 H_{\mu\nu}$ evaluated on the 0-th order background we
find
\be \mu^2 H_{00}^{(0)} \approx
-\frac{3}{4}\frac{G_Nm}{m_s^2r^5}(3-30\frac{z^2}{r^2}+35\frac{z^4}{r^4})\; , \;\;\;\;\; \mu^2
{H_\mu^\mu}^{(0)} \approx 0\,\ee
where we have defined $m_s\equiv \frac{G_NM}{r_0^3\mu}$. Taking
appropriate care in order to avoid the
  introduction of spurious sources, we can solve the
previous equations yielding
\be \Phi^{(1)}=2\Psi^{(1)} = -\frac{3}{8}\frac{G_Nm}{m_s^2 r^3}\left(1-6\frac{z^2}{r^2}+5\frac{z^4}{r^4}\right)\,.\ee
$\Phi^{(1)}$ is the first correction to the Newtonian potential in an
  expansion in powers of $1/(m_sr)$. Notice that when the correction
  to the Newtonian potential becomes of order one (for $r\sim
  m_s^{-1}$), the expansion that we have used breaks down. This was
  anticipated by our discussion above, since the method that we have
  used is only valid whenever the modification of the GR solution is
  small. But this computation is enough to show that the corrections
  are suppressed at distances bigger that $m_s^{-1}$ and they are
  anisotropic. This anisotropy is an expected property for this theory
  since, from an effective field theory point of view, the properties of
  the excitations that we have depend crucially on the background. So the
  corrections that we can expect to get from them will also reflect
  the symmetries of the background geometry. In Fig.~1 we present a
  density plot of this correction. The gravitational force is
  the gradient of this field, and it points from the darker towards
  the lighter regions. It is apparent that the attractive/repulsive
  nature of the correction depends on the direction of the measurement.

\begin{figure}
\begin{center}
\psfig{figure=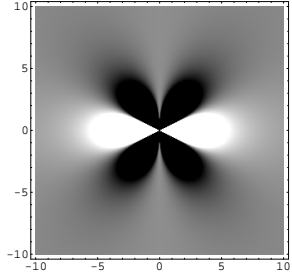,height=1.5in}
\end{center}
\caption{Density plot of the first correction to the Newtonian
  potential. Darker means higher values.
\label{fig:radish}}
\end{figure}

\section{Conclusions}

We have proposed a class of modifications of GR that have the
potential of addressing the issues of dark matter and dark energy as
manifestations of the same phenomenon. These actions naturally modify
the Newtonian potential below a fixed acceleration scale given by the
late-time Hubble constant (times the speed of light), as required in MOND. We have seen that
they are stable and ghost free and pass all tests of gravity at the
Solar System and laboratory levels. Interestingly, these models predict
gravitational anomalies at a level that is accessible in current
and planned experiments. Their gravitational phenomenology is markedly
characteristic, and to our knowledge these are the only
models that predict observable corrections to Newton's potential reflecting the
geometry of the underlying background geometry. Were such anisotropic
corrections of Newton's potential to be found, they would provide
in our view a smoking gun for the existence of a MOND-like
modification of gravity along these lines that would be responsible for
the structure observed in rotation curves of galaxies and the
acceleration of the Universe.

\section{Acknowledgements}

We thank the organizers of the XLIst Rencontres de Moriond for the invitation to give this talk.

\end{document}